\documentclass[aps,twocolumn,showpacs,preprintnumbers,superscriptaddress,nofootinbib]{revtex4}

\usepackage{subfigure,graphicx,epsfig,amsmath,amsfonts,amssymb,xcolor,slashed,ulem}

\newcommand{\be}{\begin{equation}}
\newcommand{\ee}{\end{equation}}
\newcommand{\ba}{\begin{eqnarray}}
\newcommand{\ea}{\end{eqnarray}}

\begin{document}

\title{ A meson-baryon molecular interpretation for some $\Omega_c$ excited baryons.}

\date{\today}

\author{Gl\`oria Monta\~na}
\affiliation{Departament de F\'{\i}sica Qu\`antica i Astrof\'{\i}sica and Institut de Ci\`encies del Cosmos (ICCUB), Universitat de Barcelona,
Mart\'{\i} i Franqu\`es 1, 08028 Barcelona, Spain}

\author{Albert Feijoo}
\affiliation{Departament de F\'{\i}sica Qu\`antica i Astrof\'{\i}sica and Institut de Ci\`encies del Cosmos (ICCUB), Universitat de Barcelona,
Mart\'{\i} i Franqu\`es 1, 08028 Barcelona, Spain}

\author{\`Angels Ramos}
\affiliation{Departament de F\'{\i}sica Qu\`antica i Astrof\'{\i}sica and Institut de Ci\`encies del Cosmos (ICCUB), Universitat de Barcelona,
Mart\'{\i} i Franqu\`es 1, 08028 Barcelona, Spain}


\begin{abstract} 
We explore the possibility that some of the five narrow $\Omega_c$ resonances recently observed at LHCb could correspond to pentaquark states, structured as meson-baryon bound states or molecules. The interaction of the low-lying pseudoscalar mesons with the ground-state baryons in the charm $+1$, strangeness $-2$ and isospin 0 sector is built from t-channel vector meson exchange, using effective Lagrangians. The resulting s-wave coupled-channel unitarized amplitudes show the presence of two structures with similar masses and widths to those of the observed $\Omega_c(3050)^0$ and $\Omega_c(3090)^0$. The identification of these resonances with the meson-baryon bound states found in this work would also imply assigning the values $1/2^-$ for their spin-parity. An experimental determination of the spin-parity of the $\Omega_c(3090)^0$ would help in disentangling its structure, as the quark-based models predict its spin-parity to be either $3/2^-$ or $5/2^-$.

\end{abstract}


\maketitle

\section{Introduction}

The recent observation by the LHCb collaboration of five narrow $\Omega_c^0$ excited resonances decaying into $\Xi_c^+ K^-$ states \cite{Aaij:2017nav}
has triggered a lot of activity in the field of baryon spectroscopy aiming at understanding their inner structure and possibly establishing their unknown values of spin-parity \cite{Karliner:2017kfm,Wang:2017vnc,Wang:2017zjw,Chen:2017gnu,Padmanath:2017lng,Chen:2017sci,Agaev:2017jyt,Agaev:2017lip,Cheng:2017ove,Wang:2017hej,Huang:2017dwn,Yang:2017rpg,An:2017lwg,Kim:2017jpx}.

In conventional quark models, baryons are composed by three quarks but, in spite of the rather successful description of a wealth of data \cite{Capstick:1986bm}, other more exotic components cannot be ruled out. Within a $css$ quark content picture the presumed spin-parity of the $1/2^+$ and $3/2^+$ $\Omega_c^0$ ground states \cite{pdg} can be explained naturally, and predictions for the low lying excited states have also been given \cite{Maltman:1980er,Migura:2006ep,Roberts:2007ni,Valcarce:2008dr,Ebert:2011kk,Vijande:2013yxa,Yoshida:2015tia}. The recent discovery of excited $\Omega^0$ states decaying into $K^-\Xi_c^+$ pairs at LHCb \cite{Aaij:2017nav} has provided new information against which revisited quark models can be tested. Actually, several interpretations have been proposed which, in general, benefit from the symmetries associated to the presence of a charm quark having a much larger mass than that of their strange companions. Some works interpret all the observed states as P-wave orbital excitations of the $ss$ diquark with respect to the charmed quark \cite{Karliner:2017kfm,Wang:2017vnc,Wang:2017zjw,Chen:2017gnu}, a result which finds support from a recent Lattice QCD simulation reporting the energy spectra of $\Omega_c^0$ baryons with spin up to 7/2 for both positive and negative parity \cite{Padmanath:2017lng}. In other works, some of the states would be associated to 1P orbital excitations \cite{Chen:2017sci} and others to radial 2S orbital ones\cite{Agaev:2017jyt,Agaev:2017lip,Cheng:2017ove,Wang:2017hej}. A pentaquark structure has also been advocated for the excited $\Omega_c^0$ baryons, either from a model that includes the exchange of Goldstone mesons in the interaction between the constituent quarks  \cite{Huang:2017dwn,Yang:2017rpg,An:2017lwg} or by employing the quark-soliton model \cite{Kim:2017jpx}.

An alternative scenario is provided by models that can interpret some resonances as being composed by five quarks, structured
in the form of a quasi-bound state of an interacting meson-baryon pair. A paradigmatic example is that of the $\Lambda(1405)$ resonance, the mass of which is systematically overpredicted by quark models. Instead,  dedicated studies of the meson-baryon interaction in the $I=0$ $S=-1$ sector, employing chiral effective lagrangians and implementing unitarization, predict the $\Lambda(1405)$ as being the superposition of two-poles of the meson-baryon scattering matrix \cite{pdg,Oller:2000fj,Jido:2003cb,Hyodo:2011ur}. This two-pole structure received support from the simultaneous analysis in \cite{Magas:2005vu} of different line shapes \cite{Thomas:1973uh,Prakhov:2004an} and, more recently, from the analysis of the $\pi \Sigma$ photoproduction data  \cite{Niiyama:2008rt,Moriya:2013hwg} performed in \cite{Roca:2013av,Mai:2014xna}. The existence of pentaquark baryons have been made clearly evident from the recent discovery at LHCb  \cite{Aaij:2015tga} of the excited nucleon resonances $P_c(4380)$ and $P_c(4450)$, seen in the invariant mass distribution of $J/\psi \, p$ pairs from the decay of the $\Lambda_b$. The high mass of these excited nucleons inevitably demands the presence of an additional $c\bar{c}$ pair. Hidden charm baryons having a meson-baryon structure had already been predicted  previously \cite{Wu:2010jy,Wu:2010vk,Yang:2011wz,Xiao:2013yca,Karliner:2015ina}, and later studies confirmed that the narrow pentaquark seen from the $\Lambda_b \to J/\Psi~ K^- p$ decay at CERN could receive a molecular interpretation \cite{Chen:2015loa,Roca:2015dva,He:2015cea,Meissner:2015mza}. Reactions to find the hidden charm strangeness $S=-1$ partner of the pentaquark have also been proposed recently \cite{Chen:2015sxa,Feijoo:2015kts}.

The aim of this work is to explore whether some of the observed excited $\Omega_c^0$ resonances admit an interpretation as meson-baryon molecules. 
Similarly to the $P_c$ pentaquarks, which find more natural having a $c\bar{c}$ pair in its composition rather than being an extremely high energy excitation of the $3q$ system, it is also plausible to expect that some excitations in the $C=1$, $S=-2$ sector can be obtained by adding a $u\bar{u}$ pair to the natural $ssc$ content of the $\Omega_c^0$ \cite{Yang:2017rpg}.  The hadronization of the five quarks could then lead to bound states, generated by the meson baryon interaction in coupled channels. To support this possibility, let us point out that the $\bar K\Xi_c$ and $\bar K\Xi_c^\prime$ thresholds are in the energy range of interest, and that the excited $\Omega_c^0$ baryons under study have been observed from the invariant mass of  spectrum of $K^-\Xi^+_c$ pairs.

After the successful description of the $\Lambda(1405)$ as a $\bar{K}N$ quasibound molecular state, many groups devoted efforts to find signs of compositeness in other spin, isospin and flavour sectors, and several well known states and spectral shapes of various reactions have found a more natural explanation in terms of resonances being generated by the interaction of mesons and baryons in coupled channels, see \cite{Guo:2017jvc} and references therein. 
In the particular open charm sector with strangeness $S=-2$ approached in the present work, the authors of Ref.~\cite{Hofmann:2005sw} find a rich spectrum of molecules, employing a zero-range exchange of vector mesons as the driving force for the s-wave scattering of
pseudo-scalar mesons off the baryon ground states. Similar qualitative findings where obtained in the work of Ref.~\cite{JimenezTejero:2009vq}, where finite range effects were explored. Heavy-quark spin symmetry (HQSS) is explicitly considered in the model of Ref.~\cite{Romanets:2012hm}, thus treating the pseudoscalar and vector mesons, as well as the ground state $1/2^+$ and $3/2^+$ baryons, on the same footing. In spite of their qualitative differences, the three works predict the existence of $\Omega_c^0$ resonances as poles of the coupled-channel meson baryon scattering amplitude in the complex plane. However, most of the predicted quasibound states are below 3 GeV, too low to explain the observed states. Only the model of Ref.~\cite{JimenezTejero:2009vq} predicts a state at 3117 MeV, but its width turns out to be one order of magnitude larger than that of the closer observed state.  In the present work we 
revisit the original model of Ref.~\cite{Hofmann:2005sw} and find that, after taking an appropriate regularization scheme with physically sound parameters, two $\Omega_c^0$ resonances are generated in the region of interest. Our model is able to reproduce the mass of two of the excited $\Omega_c^0$ baryons seen at CERN, at 3050 MeV and 3090~MeV, as well as their widths. The important observation is that, if these molecular states are identified with the observed ones, their spin-parity would be assigned to be $1/2^-$.

\section{Formalism}\label{sec:formalism}

\begin{figure}[htbp!]
 \centering
 		\includegraphics[width=0.45\textwidth]{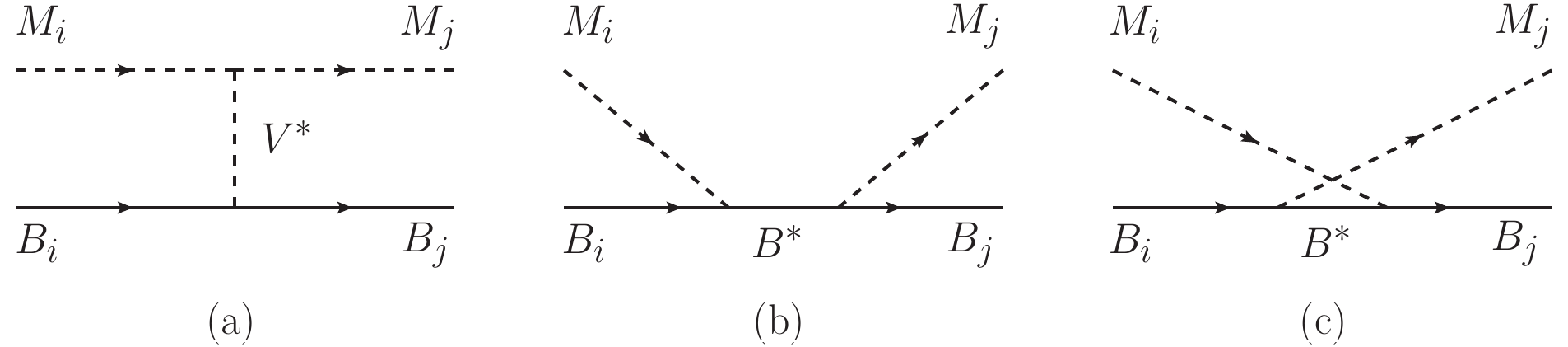}
\caption{Leading order tree level diagrams contributing to the $PB$ interaction. Baryons and pseudoscalar mesons are depicted by solid and dashed lines, respectively.}
\label{fig:feynmandiagram_ps}
\end{figure}

The diagrams contributing to the meson-baryon interaction at tree level are depicted in Fig.~\ref{fig:feynmandiagram_ps}. For the s-wave amplitude explored here, we will only consider the most important contribution which corresponds to the  t-channel  vector meson exchange term (Fig.~\ref{fig:feynmandiagram_ps}(a)). The vertices in this diagram, coupling the vector meson to pseudoscalar mesons ($VPP$) and baryons ($VBB$), 
are described from effective Lagrangians, that are obtained using the hidden gauge formalism and assuming $SU(4)$ symmetry  \cite{Hofmann:2005sw}:
\begin{equation}\label{eq:vertxVPP}
\mathcal{L}_{VPP}=ig\langle\left[\partial_\mu\phi, \phi\right] V^\mu\rangle,
\end{equation}
\begin{equation}\label{eq:vertexBBV}
\mathcal{L}_{VBB}=\frac{g}{2}\sum_{i,j,k,l=1}^4\bar{B}_{ijk}\gamma^\mu\left(V_{\mu,l}^{k}B^{ijl}+2V_{\mu,l}^{j}B^{ilk}\right),
\end{equation}
where the symbol $\langle~\rangle$ denotes the trace of $SU(4)$ matrices in flavour space, and the factor $g$ is the universal coupling constant, related to the pion decay constant $f=93\rm~MeV$ by:
\begin{equation}\label{eq:g_coup}
g=\frac{m_V}{2f},
\end{equation}
with $m_V$ being a representative mass of the light (uncharmed) vector mesons from the nonet. This value of $g$ is in accordance with the KSFR\footnote{Kawarabayashi-Suzuki-Fayyazuddin-Riazuddin} relation \cite{Kawarabayashi:1966kd,Riazuddin:1966sw}.

The symbols $\phi$ and $V_\mu$ represent the pseudoscalar fields of the 16-plet of the $\pi$ and the vector fields of the 16-plet of the $\rho$, 
given by
\begin{equation}\label{eq:matrixphi}
{\scriptsize\phi \! = \!
\begin{pmatrix}
\frac{1}{\sqrt{2}}\pi^0+\frac{1}{\sqrt{6}}\eta+\frac{1}{\sqrt{3}}\eta^\prime & \pi^+ & K^{+} &\ \bar{D}^{0} \\
\pi^- & \!\!\!\!\!\!\! -\frac{1}{\sqrt{2}}\pi^0+\frac{1}{\sqrt{6}}\eta+\frac{1}{\sqrt{3}}\eta^\prime & K^{0} & D^{-} \\
K^{-} & \bar{K}^{0} & \!\!\!\!\!\!\!-\sqrt{\frac{2}{3}}\eta+\frac{1}{\sqrt{3}}\eta^\prime & D_s^{-} \\
D^{0} & D^{+} & D_s^{+} & \eta_c \\
\end{pmatrix}},
\end{equation}
and
\begin{equation}\label{eq:matrixVmu}
{\scriptsize V_\mu =
\begin{pmatrix}
\frac{1}{\sqrt{2}}(\rho^0+\omega) & \rho^+ & K^{\ast +} & \bar{D}^{\ast 0} \\
\rho^- & \frac{1}{\sqrt{2}}(-\rho^0+\omega) & K^{\ast 0} & D^{\ast -} \\
K^{\ast -} & \bar{K}^{\ast 0} & \phi & D_s^{\ast -} \\
D^{\ast 0} & D^{\ast +} & D_s^{\ast +} & J/\psi \\
\end{pmatrix}_\mu,}
\end{equation}
respectively, and $B$ is the tensor of baryons belonging to the 20-plet of the $p$:
\begin{equation}
{\scriptsize\begin{tabular}{ll}
$B^{121}=p$, & ~~~~~$B^{122}=n$, 
$\vphantom{\sqrt{\frac{2}{3}}}$ \\
 $B^{132}=\frac{1}{\sqrt{2}}\Sigma^0-\frac{1}{\sqrt{6}}\Lambda$,  & ~~~~~$B^{213}=\sqrt{\frac{2}{3}}\Lambda$,  
 $\vphantom{\sqrt{\frac{2}{3}}}$\\
 $B^{231}=\frac{1}{\sqrt{2}}\Sigma^0+\frac{1}{\sqrt{6}}\Lambda$, & ~~~~~$B^{232}=\Sigma^-$, 
 $\vphantom{\sqrt{\frac{2}{3}}}$\\
$B^{233}=\Xi^-$, & ~~~~~$B^{311}=\Sigma^+$, 
$\vphantom{\sqrt{\frac{2}{3}}}$\\
$B^{313}=\Xi^0$, & ~~~~~$B^{141}=-\Sigma_c^{++}$, 
 $\vphantom{\sqrt{\frac{2}{3}}}$\\
$B^{142}=\frac{1}{\sqrt{2}}\Sigma_c^++\frac{1}{\sqrt{6}}\Lambda_c$, & ~~~~~$B^{143}=\frac{1}{\sqrt{2}}\Xi_c^{'+}-\frac{1}{\sqrt{6}}\Xi_c^+$, 
$\vphantom{\sqrt{\frac{2}{3}}}$\\
$B^{241}=\frac{1}{\sqrt{2}}\Sigma_c^+-\frac{1}{\sqrt{6}}\Lambda_c$, & ~~~~~$B^{242}=\Sigma_c^0$, 
$\vphantom{\sqrt{\frac{2}{3}}}$\\
$B^{243}=\frac{1}{\sqrt{2}}\Xi_c^{'0}+\frac{1}{\sqrt{6}}\Xi_c^0$, & ~~~~~$B^{341}=\frac{1}{\sqrt{2}}\Xi_c^{'+}+\frac{1}{\sqrt{6}}\Xi_c^+$, 
$\vphantom{\sqrt{\frac{2}{3}}}$\\
$B^{342}=\frac{1}{\sqrt{2}}\Xi_c^{'0}-\frac{1}{\sqrt{6}}\Xi_c^0$, & ~~~~~$B^{343}=\Omega_c^0$, 
$\vphantom{\sqrt{\frac{2}{3}}}$\\
$B^{124}=\sqrt{\frac{2}{3}}\Lambda_c$, & ~~~~~$B^{234}=\sqrt{\frac{2}{3}}\Xi_c^0$, 
$\vphantom{\sqrt{\frac{2}{3}}}$\\ 
$B^{314}=\sqrt{\frac{2}{3}}\Xi_c^+$, & ~~~~~$B^{144}=\Xi_{cc}^{++}$, 
$\vphantom{\sqrt{\frac{2}{3}}}$\\ 
$B^{244}=-\Xi_{cc}^+$, & ~~~~~$B^{344}=\Omega_{cc}$, 
$\vphantom{\sqrt{\frac{2}{3}}}$\\ 
\end{tabular}}
\end{equation}
where the indices $i,j,k$ of $B^{ijk}$ denote the quark content of the baryon fields with the identification $1\leftrightarrow u$, $2\leftrightarrow d$, $3\leftrightarrow s$ and $4\leftrightarrow c$. 
The phase convention for the isospin states is $\mid{\pi^+}\rangle=-\mid{1,1}\rangle$, $\mid{K^{ -}}\rangle=-\mid{1/2,-1/2}\rangle$ and $\mid{D^{ 0}}\rangle=-\mid{1/2,-1/2}\rangle$ for the pseudoscalar mesons and, analogously,  $\mid{\rho^+}\rangle=-\mid{1,1}\rangle$, $\mid{K^{*\, -}}\rangle=-\mid{1/2,-1/2}\rangle$ and $\mid{D^{*\, 0}}\rangle=-\mid{1/2,-1/2}\rangle$ for the vector mesons. For the baryons, we take $\mid{\Sigma^+}\rangle=-\mid{1,1}\rangle$ and $\mid{\Xi^-}\rangle=-\mid{1/2,-1/2}\rangle$. This convention is consistent with the structure of the $\phi$, $V_\mu$ and $B$ fields. It is the one followed in Refs. \cite{Wu:2010jy,Gamermann:2006nm} and it differs from that  in Ref.~\cite{Hofmann:2005sw} in the sign of the $D^+(D^{\ast +})$ and $D^-(D^{\ast -})$ mesons.

Using the $VPP$ and $VBB$ vertices above one obtains the t-channel Vector-Meson-Exchange (TVME) potential \cite{Hofmann:2005sw}:
\begin{eqnarray}\label{eq:Vij_1}
 V_{ij}&=&g^2 \sum_v C_{ij}^v \bar{u}\left(p_j\right)\gamma^\mu u\left(p_i\right)\frac{1}{t-m_v^2} \nonumber \\
 & & \phantom{g^2 \sum_v } \times \left[\left(k_i+k_j\right)_\mu -\frac{k_i^2-k_j^2}{m_v^2}\left(k_i-k_j\right)_\mu\right],
\end{eqnarray}
where $p_i$, $p_j$ ($k_i$, $k_j$) are the four-momenta of the baryons (mesons) in the $i$, $j$ channels and $m_v$ is the vector meson mass. Adopting the same mass $m_v=m_V$ for the light vector mesons and accounting for the higher mass of the charmed mesons with a common multiplying factor $\kappa_c=(m_V/m_V^c)^2\approx 1/4$ as in \cite{Mizutani:2006vq}, Eq.~(\ref{eq:Vij_1}) simplifies to
\begin{equation}\label{eq:Vij_2}
 V_{ij}=-C_{ij}\frac{1}{4f^2}\bar{u}\left(p_j\right)\gamma^\mu u\left(p_i\right)\left(k_i+k_j\right)_\mu,
\end{equation}
where the limit  $t\ll m_V$ has been taken to reduce the t-channel diagram to a contact term. The coefficients $C_{ij}$  are symmetric with respect to the indices and are obtained summing the various vector meson exchange contributions, $ \sum_v C_{ij}^v$  \cite{Hofmann:2005sw}, including the factor $\kappa_c$ in the case of charmed mesons. Working out the Dirac algebra up to order ${\cal O}(p^2/M)$ corrections, one gets the expression:
\begin{equation}\label{eq:Vij}
 V_{ij}(\sqrt{s})=-C_{ij}\frac{1}{4f^2}\left(2\sqrt{s}-M_i-M_j\right) N_i N_j
\end{equation}
where $M_i$, $M_j$ and $E_i$, $E_j$ are the masses and the energies of the baryons and $N= 
 \sqrt{(E+M)/2M}$. Note that, while $SU(4)$ symmetry is encoded in the values of the coefficients $C^v_{ij}$, the interaction potential is not SU(4) symmetric due to the use of physical masses for the mesons and baryons involved, as well as to the factor $\kappa_c$.

For the isospin $I=0$, charm $C=1$ and strangeness $S=-2$ sector studied here, the available pseudoscalar-baryon channels are $\bar{K}\Xi_c (2964)$, $\bar{K}\Xi'_c (3070)$, $D\Xi (3189)$, $\eta \Omega_c (3246)$, $\eta' \Omega_c (3656)$, $\bar{D}_s \Omega_{cc} (5528)$, and $\eta_c \Omega_c (5678)$, where the values in parentheses indicate their corresponding threshold. The doubly charmed $\bar{D}_s \Omega_{cc}$ and $\eta_c \Omega_c $ channels will be neglected, as their energy is much larger than that of the other channels. We have checked that their inclusion barely influences the results presented here. The matrix of $C_{ij}$ coefficients for the resulting 5-channel interaction is given in Table~\ref{tab:coeff}.

\begin{table}[h]
\begin{tabular}{l c c c c c}
\hline
 & & & & &  \\
 [-2.5mm]
  &{${\bar K}\Xi_c$}  & {${\bar K}\Xi_c^\prime$}  & { $D\Xi$}  & { $\eta\Omega_c^0$} &{$\eta^\prime\Omega_c^0$}  \\
\hline
 & & & & &  \\[-2.5mm]
{${\bar K}\Xi_c$}         & $1$ & $0$ & $\sqrt{\displaystyle\frac{3}{2}}\kappa_c$  & $0$ & $0$     \\
{${\bar K}\Xi_c^\prime$}   &     & $1$ & $\displaystyle\frac{1}{\sqrt{2}}\kappa_c$ & $-\sqrt{6}$ & $0$   \\
{ $D\Xi$} &     &     &    $2$        &  $-\displaystyle\frac{1}{\sqrt{3}}\kappa_c$  & $-\sqrt{\displaystyle\frac{2}{3}}\kappa_c$ \\
{ $\eta\Omega_c^0$} &     &     &     &  $0$  &  $0$\\
{ $\eta^\prime\Omega_c^0$}  &     &     &    &     &  $0$   \\

\hline

\end{tabular}
\caption{The $C_{ij}$ coefficients for the $I=0$, $C=1$, $S=-2$ sector of the  $PB$ interaction.}
\label{tab:coeff}
\end{table}

The interaction of vector mesons with baryons is obtained following the formalism presented in Ref.~\cite{Oset:2009vf}, which is extended to SU(4) here. Similarly as for pseudoscalar mesons, we only retain the t-channel vector-exchange term.  Employing the effective Lagrangian:
\begin{equation}\label{eq:vertexVVV}
\mathcal{L}_{VVV}=ig\langle {\left[V^\mu,\partial_\nu V_\mu\right] V^\nu}\rangle .
\end{equation}
for the three-vector $VVV$ vertex and that of Eq.~(\ref{eq:vertexBBV}) for the $VBB$ one, the resulting interaction kernel for the vector-baryon ($VB$) interaction is identical to that obtained for the pseudoscalar-baryon ($PB$) one (see Eq.~(\ref{eq:Vij})), multiplied by the product of polarization vectors, $\vec{\epsilon}_i,\;\vec{\epsilon}_j$.

The allowed vector meson-baryon states are 
$D^*\Xi (3326)$, $\bar{K}^*\Xi_c (3363)$, $\bar{K}^*\Xi'_c (3470)$, $\omega \Omega_c (3480)$, $\phi \Omega_c (3717)$, $\bar{D}_s^* \Omega_{cc} (5662)$ and $J/\psi \Omega_c (5794)$ , where, again, we will neglect the doubly charmed states. The coefficients $C_{ij}$ can be straightforwardly obtained from those for the $PB$ interaction in Table~\ref{tab:coeff}, by considering the following correspondences: 
\begin{equation}
 \begin{matrix}
  \pi \rightarrow \rho, & K\rightarrow K^\ast, & \bar{K}\rightarrow\bar{K}^\ast, & D\rightarrow D^\ast, & \bar{D}\rightarrow\bar{D}^\ast, 
 \end{matrix}
\end{equation}
\begin{equation}
 \begin{matrix}
  \frac{1}{\sqrt{3}}\eta+\sqrt{\frac{2}{3}}\eta^\prime\rightarrow\omega {\qquad\rm and\;} & -\sqrt{\frac{2}{3}}\eta+\frac{1}{\sqrt{3}}\eta^\prime\rightarrow\phi\ .
 \end{matrix}
 \label{eq:eta_phi}
\end{equation}

\begin{figure}[htbp!]
 \centering
	\includegraphics[width=0.5\textwidth]{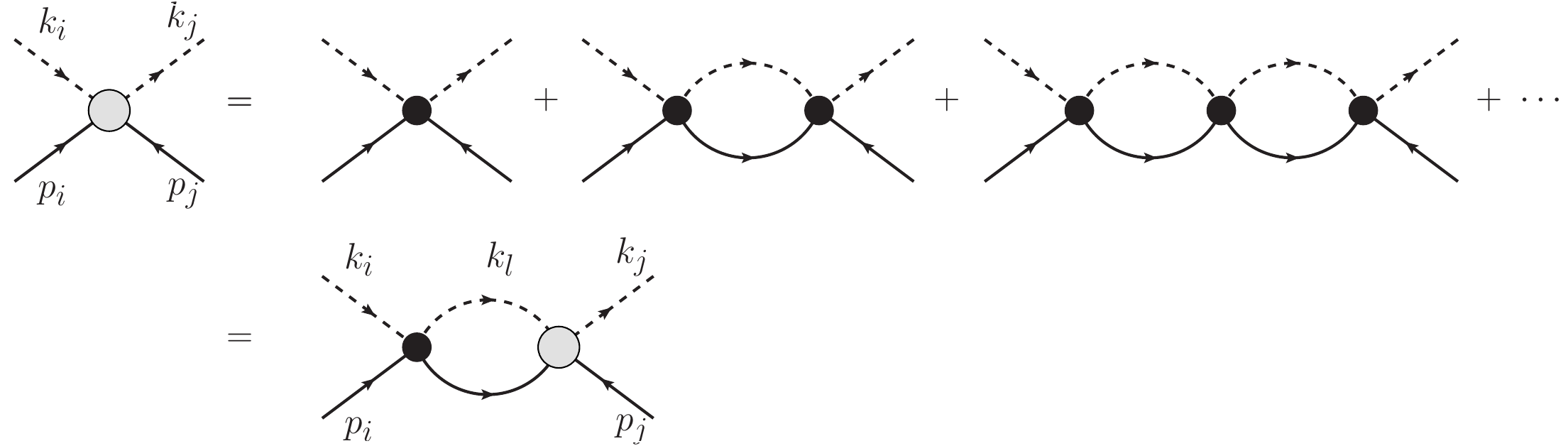}
\caption{Diagrams representing the Bethe-Salpeter equations in meson-baryon ($MB$) scattering. The big empty circle corresponds to the $T_{ij}$ matrix element, the black circles correspond to the potential $V_{ij}$ and the loops represent the propagator $G_{l}$ function. The $i, j, l$ indices stand for the channels of the coupled-channel theory.}
\label{fig:BSfeynman}
\end{figure}

The sought resonances will be generated as poles of the scattering amplitude $T_{ij}$, unitarized via the coupled-channel Bethe-Salpeter (B-S) equation, which implements the resummation of loop diagrams to infinite order schematically depicted in Fig.~\ref{fig:BSfeynman} and has the expression
\begin{equation}\label{eq:BSeq}
T_{ij}=V_{ij}+V_{il}G_{l}T_{lj}.
\end{equation}
Factorizing the $V$ and $T$ matrices on-shell out of the internal integrals, the solution of the former equation
\begin{equation}\label{eq:BSeq2}
T=(1-VG)^{-1}V
\end{equation}
is purely algebraic.  
We note that the sum over the polarizations of the internal vector mesons gives
\begin{equation}
 \sum_{\rm pol}\epsilon_i\epsilon_j=\delta_{ij}+\frac{q_iq_j}{M_V^2} \ ,
\end{equation}
and, neglecting the correction $\sim q^2/M_V^2$, which is consistent with the approximations done so far, 
the factor $\vec{\epsilon}_i\vec{\epsilon}_j$ can be factorized out in all the terms of the B-S equation.

The loop function is given by
 \begin{equation}\label{eq.Gmatrix}
G_{l}=i\int \frac{d^4q}{(2\pi)^4}\frac{2M_l}{(P-q)^2-M_l^2+i\epsilon}\frac{1}{q^2-m_l^2+i\epsilon},
\end{equation}
where $M_l$ and $E_l$ correspond to the mass and the energy of the intermediate baryon, $m_l$ is the mass of the intermediate meson, $P=k+p=(\sqrt{s},\vec{0})$ is the total four-momentum of the system in the c.m. frame and $q$ denotes the four-momentum of the meson propagating in the intermediate loop.
This function diverges for $\vec{q}\rightarrow\infty$ and it must be regularized with a proper scheme. One may employ the  \textit{cut-off} regularization method, which consists in replacing the infinite upper limit of the integral by a large enough cut-off momentum $\Lambda$,
\begin{equation}\label{eq:determine_qcut}
 G_{l}^{\rm cut}=\int_{0}^{\Lambda}\frac{d^3q}{(2\pi)^3}\frac{1}{2\omega_l(\vec{q})}\frac{M_l}{E_l(\vec{q})}\frac{1}{\sqrt{s}-\omega_l(\vec{q})-E_l(\vec{q})+i\epsilon},
\end{equation}
or the alternative \textit{dimensional regularization} (DR) approach, which is the one adopted here:
\begin{equation}\label{eq:GmatrixDR}
\begin{aligned}
 G_{l}=&\frac{2M_l}{16\pi^2}\Big\{ a_l(\mu)+\ln\frac{M_l^2}{\mu^2}+\frac{m_l^2-M_l^2+s}{2s}\ln\frac{m_l^2}{M_l^2}+ \\ 
  &+\frac{q_l}{\sqrt{s}}\left[\ln\left(s-(M_l^2-m_l^2\right)+2q_l\sqrt{s})\right. \\
  &\phantom{\frac{q_l}{\sqrt{s}}~~}+\ln\left(s+(M_l^2-m_l^2\right)+2q_l\sqrt{s})\\
  &\phantom{\frac{q_l}{\sqrt{s}}~~}-\ln\left(-s+(M_l^2-m_l^2\right)+2q_l\sqrt{s})\\
  &\phantom{\frac{q_l}{\sqrt{s}}~~}\left.-\ln\left(-s-(M_l^2-m_l^2\right)+2q_l\sqrt{s}) \right] \Big\},
\end{aligned}
\end{equation}
where $a_l(\mu)$ is the subtraction constant at the regularization scale $\mu$, and $q_l$ is the on-shell three-momentum of the meson in the loop.
The choice of the regularization scale $\mu$ and the corresponding subtraction constants $a_l(\mu)$ can be obtained by demanding that, at an energy close to the channel threshold, $G_l$ is similar to $G_l^\text{cut}$ for a certain cut-off $\Lambda$, namely
\begin{equation}\label{eq:a(mu)}
 a_l({\mu})= \frac{16\pi^2}{2M_l}\left(G_{l}^\text{cut}(\Lambda)-G_{l}(\mu,a_l=0)\right).
\end{equation}
The value of $\Lambda$ is usually taken around several hundreds of MeV, which is around the scale that has been integrated out in the zero range approximation of the meson-exchange model considered here.  Typical values of the DR parameters are $\mu\approx630\rm~MeV$ and $a(\mu)\sim-2.0$ in the case of $SU(3)$ (see Ref.~\cite{Oller:2000fj}) while in $SU(4)$ previous works have taken $\mu=1000\rm~MeV$ and $a(\mu)\sim-2.3$ \cite{Wu:2010jy,Wu:2010vk}.

The expression for the loop function $G_l$ in Eq.~(\ref{eq:GmatrixDR}) assumes that the baryon and the meson have fixed masses and no width.
When the B-S equation involves channels that include particles with a large width, which is the case of the $\rho$ ($\Gamma_\rho=149.4\rm~MeV$) and $K^\ast$ ($\Gamma_{K^\ast}~=~50.5\rm~MeV$) mesons, this function has to be convoluted with the mass distribution of the particle. Following the method described in \cite{Oset:2009vf}, the loop function in these cases will be replaced by
\begin{equation}
\begin{aligned}
 \tilde{G}_{l}(s)=-\frac{1}{N}\int_{(m_l-2\Gamma_l)^2}^{(m_l+2\Gamma_l)^2} & \frac{d\tilde{m}_l^2}{\pi}{\rm\, Im\,}\frac{1}{\tilde{m}_l^2-m_l^2+i\,m_l\Gamma(\tilde{m}_l)} \\
 & \times G_{l}\left(s,\tilde{m}_l^2,M_l^2\right),
\end{aligned}
\end{equation}
where we have taken the limits of the integral to extend over a couple of times the width of the meson, and the normalization factor $N$ reads
\begin{equation}
 N=\int_{(m_l-2\Gamma_l)^2}^{(m_l+2\Gamma_l)^2}d\tilde{m}_l^2\left(-\frac{1}{\pi}\right){\rm\, Im\,}\frac{1}{\tilde{m}_l^2-m_l^2+i\,m_l\Gamma(\tilde{m}_l)} \ .
\end{equation}
The energy dependent width is given by
\begin{equation}\label{eq:G_kallen}
 \Gamma(\tilde{m}_l)=\Gamma_l\frac{m_l^5}{\tilde{m}_l^5}\frac{\lambda^{3/2}(\tilde{m}_l^2,m_1^2,m_2^2)}{\lambda^{3/2}(m_l^2,m_1^2,m_2^2)}\,\theta(\tilde{m}_l-m_1-m_2),
\end{equation}
where
$m_1$ and $m_2$ are the masses of the lighter mesons to which the vector meson in the loop decays, i.e. $m_1=m_2=m_\pi$ for the $\rho$ and $m_1=m_\pi$, $m_2=m_K$ for the $K^\ast$ and  $\lambda$ is the K\"{a}ll\'{e}n function $\lambda(x,y,z)=(x-(\sqrt{y}+\sqrt{z})^2)(x-(\sqrt{y}-\sqrt{z})^2)$. 

A resonance generated dynamically from the coupled channel meson-baryon interaction appears as a pole of the scattering amplitude $T$ in the so-called {\it second Riemann sheet} (${\rm II}$) of the complex energy plane, which implies performing the calculation of the loop function given in Eq.~(\ref{eq:GmatrixDR}) with a rotated momentum ($q_l\to -q_l$) or, equivalently, employing  \cite{Roca:2005nm}
\begin{equation}
 G_{l}^\text{II}(s)=G_{l}(s)+i\,2M_l\frac{q_l}{4\pi\sqrt{s}}.
\end{equation}
In the case of multiple channels, the loop function of each channel is rotated to the second Riemann sheet only when the real part of the complex energy $z\equiv \sqrt{s}$ is larger than the corresponding channel threshold. In the vicinity of a pole, $z_p$, one may write 
\begin{equation}\label{eq:pole}
T_{ij}(s)\sim \frac{g_i g_j} {z-z_p} \ ,
\end{equation}
and the coupling constants of the resonance to the various channels are obtained from the corresponding residues, calculated from:
 \begin{equation}\label{eq:coup_der}
g_i g_j=\left[\frac{\partial}{\partial z}\left.\left(\frac{1}{T_{ij}(z)}\right)\right|_{z_p}\right]^{-1} \ .
 \end{equation}
 

The amount of $i^{\rm th}$-channel meson-baryon component in a given resonance can be obtained from the real part of:
\begin{equation}\label{eq:coup_der}
X_i = - g_i^2\left.\left(\frac{\partial{G}}{\partial(\sqrt{s})}\right)\right|_{z_p} \ .
\end{equation}
This expression is based on the model-independent relation between the compositeness of a weakly bound state
and the threshold parameters of the interaction generating it,  derived in Ref.~\cite{Weinberg:1965zz} . This idea has been reformulated within a field theoretical approach and extended to higher partial waves 
as well as to unstable (resonance) states \cite{Gamermann:2009uq,Hyodo:2011qc,Aceti:2012dd,Hyodo:2013nka,Aceti:2014ala}.

\section{Results}\label{sec:results}

In this section we present  the results obtained employing the unitarized model for meson-baryon scattering in coupled channels described above. We first describe the results obtained with the pseudoscalar-baryon interaction kernel of  Eq.~(\ref{eq:Vij}), employing the subtraction constants listed under ``Model 1" of Table~\ref{tab:a_pseudo} in the loop functions. These subtraction constants are obtained for a regularization scale of $\mu=1000$~GeV and imposing the loop function of each pseudoscalar-baryon channel to coincide, at the corresponding threshold, with the cut-off loop function evaluated for $\Lambda=800$~MeV, see Eq.~(\ref{eq:a(mu)}). We assume this value to be a natural choice as it roughly corresponds to the mass of the exchanged vector mesons in the t-channel diagram that has been eliminated in favor of the contact interaction employed in this work.
In this case, the scattering amplitude $T$ shows two poles having the following properties:
$$M_1 =  {\rm Re}\,z_1= 3051.6~{\rm MeV},~~\Gamma_1 = -2 {\rm Im} \,z_1= 0.45~{\rm MeV}$$
and
$$M_2 =  {\rm Re} \,z_2 = 3103.3~{\rm MeV},~~\Gamma_2 = -2 {\rm Im} \,z_1= 17~{\rm MeV} .$$
These resonances have spin-parity $J^P=1/2^-$, as they are obtained from the scattering amplitude of pseudoscalar mesons with baryons of the ground state octet in s-wave. 
The couplings of each resonance to the various meson-baryon channels are displayed in Table~\ref{tab:pseudo} under the label ``Model 1", where one can also find the corresponding compositeness, given by  Eq.~(\ref{eq:coup_der}), which measures the amount of each meson-baryon component in the resonance. We observe that the lowest energy state at 3052 MeV couples appreciably to the channels $\bar{K}\Xi'_c$, $D\Xi $ and  $\eta \Omega_c^0$. Note that, although the coupling to  $\eta \Omega_c^0$ states is the strongest, the compositeness is larger in the $\bar{K}\Xi'_c$ channel, to which the resonance also couples strongly and, in addition, lies closer to the corresponding threshold. The higher energy resonance at 3103 MeV, with a strong coupling to $D\Xi$ and a compositeness in this channel of 0.90, clearly qualifies as being a $D\Xi $ bound state.

\begin{table}[h]
\centering
\begin{tabular}{lccccc}
\hline 
\hline 
    &  $a_{ \bar{K}\Xi_c}$  &  $a_{\bar{K}\Xi'_c}$  &  $a_{D\Xi }$  &  $a_{\eta \Omega_c}$  & $a_{\eta' \Omega_c }$ \\
\hline \\[-2.5mm]                             
   Model 1     &  $-2.19$  &  $-2.26$  &  $-1.90$  & $-2.31$  &  $-2.26$  \\
   $\Lambda$ (MeV) & 800 & 800 & 800 & 800 & 800 \\
   \hline
    Model 2     &  $-1.69$  &  $-2.09$  &  $-1.93$ &  $-2.46$  &  $-2.42$  \\
       $\Lambda$ (MeV) & 320 & 620 & 830 & 980 & 980 \\
\hline
\hline
\end{tabular}
\caption{{\small  Values of the subtraction constants at a regularization scale $\mu=1$ GeV and the equivalent cut-off $\Lambda$ for the two models discussed in this work.}} 
\label{tab:a_pseudo}
\end{table}

\begin{table}[h!]
\centering
\begin{tabular}{c|cc|cc}
\hline
\hline
\multicolumn{5}{c}{ {\bf $0^- \oplus \frac{1}{2}^+$} interaction in {\bf$(I,S,C)=(0,-2,1)$} sector } \\ 
\hline \\[-2.5mm]
  &  \multicolumn{4}{c}{Model 1}   \\ 
\hline \\[-2.5mm]
$M\;\rm[MeV]$             &    \multicolumn{2}{c|}{$3051.6$}    &   \multicolumn{2}{c}{$3103.3$}  \\
$\Gamma\;\rm[MeV]$   &     \multicolumn{2}{c|}{$0.45$}         &     \multicolumn{2}{c}{$17$}   \\ \hline \\[-2.5mm]
                                &   $| g_i|$   & $-g_i^2 dG/dE$        &   $| g_i|$   & $-g_i^2 dG/dE$     \\
$\bar{K}\Xi_c (2964)$   &  $0.11$  & $0.00+i\,0.00$    &   $0.58$  & $0.01+i\,0.03$      \\
$\bar{K}\Xi'_c (3070)$  &  $1.67$  & $0.54+i\,0.01$    &   $0.30$  & $0.01-i\,0.01$      \\
$D\Xi (3189)$           &  $1.10$  & $0.05-i\,0.01$    &   $4.08$  & $0.90-i\,0.05$      \\
$\eta \Omega_c (3246)$  &  $2.08$  & $0.23+i\,0.00$    &   $0.44$  & $0.01+i\,0.01$      \\
$\eta' \Omega_c (3656)$ &  $0.04$  & $0.00+i\,0.00$    &   $0.28$  & $0.00+i\,0.00$      \\
\hline
\hline
 &  \multicolumn{4}{c}{Model 2} \\ 
\hline \\[-2.5mm]
$M\;\rm[MeV]$             &    \multicolumn{2}{c|}{$3050.3$}    &    \multicolumn{2}{c}{$3090.8$}   \\
$\Gamma\;\rm[MeV]$   &     \multicolumn{2}{c|}{$0.44$}        &    \multicolumn{2}{c}{$12$}     \\ \hline \\[-2.5mm]
                                &      $| g_i|$    & $-g_i^2 dG/dE$    &   $| g_i|$  & $-g_i^2 dG/dE$ \\
$\bar{K}\Xi_c (2964)$   &  $0.11$  & $0.00+i\,0.00$     &    $0.49$  & $-0.02+i\,0.01$  \\
$\bar{K}\Xi'_c (3070)$  &  $1.80$  & $0.61+i\,0.01$     &    $0.35$  & $0.02-i\,0.02$  \\
$D\Xi (3189)$           &  $1.36$  & $0.07-i\,0.01$     &    $4.28$  & $0.91-i\,0.01$ \\
$\eta \Omega_c (3246)$  &  $1.63$  & $0.14+i\,0.00$     &    $0.39$  & $0.01+i\,0.01$ \\
$\eta' \Omega_c (3656)$ &  $0.06$  & $0.00+i\,0.00$     &    $0.28$  & $0.00+i\,0.00$ \\
\hline
\hline
\end{tabular}
\caption{
{\small  $\Omega^0_c (X)$ states generated employing vector-type Weinberg-Tomozawa zero-range interactions (Model 1, Model 2) between a pseudoscalar meson and a ground state baryon, within a coupled-channel approach.}}
\label{tab:pseudo}
\end{table}

We see that our two resonances have energies very similar to the second and fourth $\Omega_c^0$ states discovered by LHCb \cite{Aaij:2017nav},  with properties:
\begin{eqnarray}
\Omega_c(3050)^0:~~& M=3050.2\pm0.1\pm0.1^{+0.3}_{-0.5}~{\rm MeV}, \nonumber \\
                                 &\Gamma=0.8\pm0.2\pm0.1~{\rm MeV},\nonumber \\
\Omega_c(3090)^0:~~& M=3090.2\pm0.3\pm0.5^{+0.3}_{-0.5}~{\rm MeV}, \nonumber \\
                                 &\Gamma=8.7\pm1.0\pm0.8~{\rm MeV}.
\label{eq:exp}
 \end{eqnarray}
We note that, even if the mass of our heavier state is larger by 10 MeV and its width is about twice the experimental one, our results clearly show the ability of the meson baryon dynamical models for generating states in the energy range of interest.

\begin{figure}[htb]
\centering
  \includegraphics[width=0.50\textwidth]{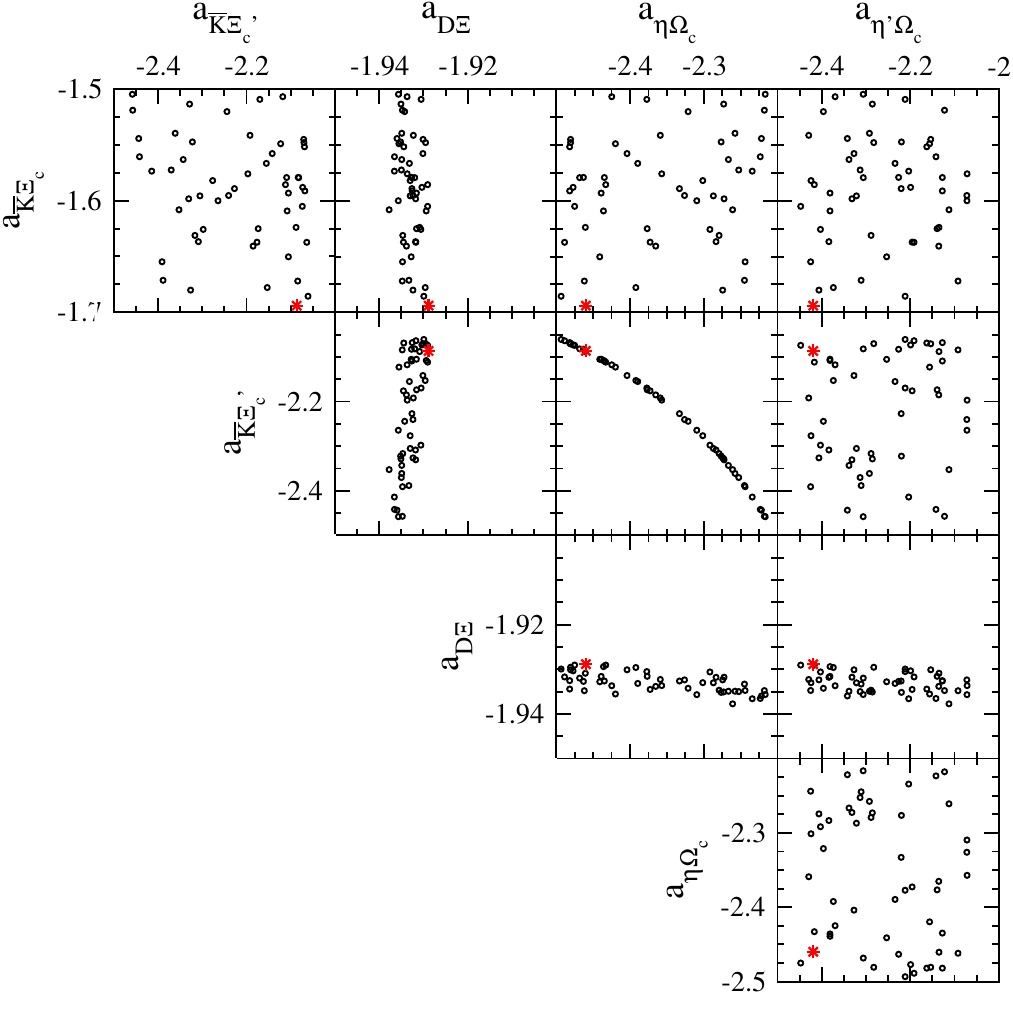}
\caption{(Color online) Correlations between the various subtraction constants. The circles represent different configurations of subtraction constants that reproduce the experimental resonances $\Omega_c(3050)^0$ and $\Omega_c(3090)^0$. The red asterisks denote one particular representative set.  } 
  \label{fig:a_corr}
\end{figure}

In an attempt to accommodate better to the data, we relax the condition of forcing that each loop function matches, at the corresponding threshold, the cut-off loop function evaluated for a cut-off $\Lambda=800$~MeV.
To this end, we let the values of the five subtracting constants vary freely within a reasonably constrained range and look for sets that reproduce the characteristics of the two observed states, $\Omega_c(3050)^0$
and $\Omega_c(3090)^0$, within $2\sigma$ of the experimental errors [see Eq.~(\ref{eq:exp})]. In order to analyze the correlations, we represent in Fig.~\ref{fig:a_corr} the values of each subtraction constant against all the others in the sets that comply with the experimental constraints. We clearly observe an anti-correlation between the subtraction constants $a_{\bar{K}\Xi'_c}$ and $a_{\eta \Omega_c}$. This can be simply understood by noting that the resonance at 3050~MeV couples mostly to these two meson-baryon states, as can be seen from the results in Table \ref{tab:pseudo}, implying that, if one subtraction constant becomes more negative, favoring a stronger attraction for the pole, the other subtraction constant needs to compensate this effect by being less negative. We also find the subtraction constant $a_{D\Xi}$ to acquire a rather stable value between -1.94 and -1.93. This is clearly a reflection of the resonance at  3090 MeV being essentially a ${D\Xi}$ bound state, which requires a particular value of the subtraction constant $a_{D\Xi}$ to generate the pole at the appropriate experimental energy. 

Among all the possible configurations of subtraction constants producing the experimental $\Omega_c^0$ states at 3050~MeV and 3090~MeV represented in Fig.~\ref{fig:a_corr}, we select a representative set, denoted by red asterisks in the figure, the values of which are listed in Table \ref{tab:a_pseudo} under the label ``Model 2". 
The two poles of the scattering amplitude of ``Model 2" have the properties:
$$M_1 =  {\rm Re}\, z_1 = 3050.3~{\rm MeV},~~~\Gamma_1 = -2 {\rm Im} \,z_1= 0.44~{\rm MeV}$$
$$M_2 =  {\rm Re} \,z_2 = 3090.8~{\rm MeV},~~~\Gamma_2 = -2 {\rm Im}\, z_1= 12~{\rm MeV} \ ,$$
which are similar for any of the sets of subtracting constants represented in Fig.~\ref{fig:a_corr}. As we see, the stronger changes are found in the higher resonance, which, apart from having been lowered to the experimental energy,  its width has been substantially decreased to agree with the experiment at $2\sigma$ level.
We see from Table \ref{tab:a_pseudo} that the equivalent values of the cut-off for this new set of subtracting constants now lie
in the range $[320-950]\;\rm MeV$. Note that the strongest change corresponds to the subtraction constant $a_{\bar{K}\Xi_c}$ , needed to decrease the width of the $\Omega_c(3090)^0$ towards its experimental value. The equivalent cut-off value of 320 MeV is on the low side of the usually employed values but it still naturally sized.

\begin{figure}[htb]
\centering
  \includegraphics[width=0.45\textwidth]{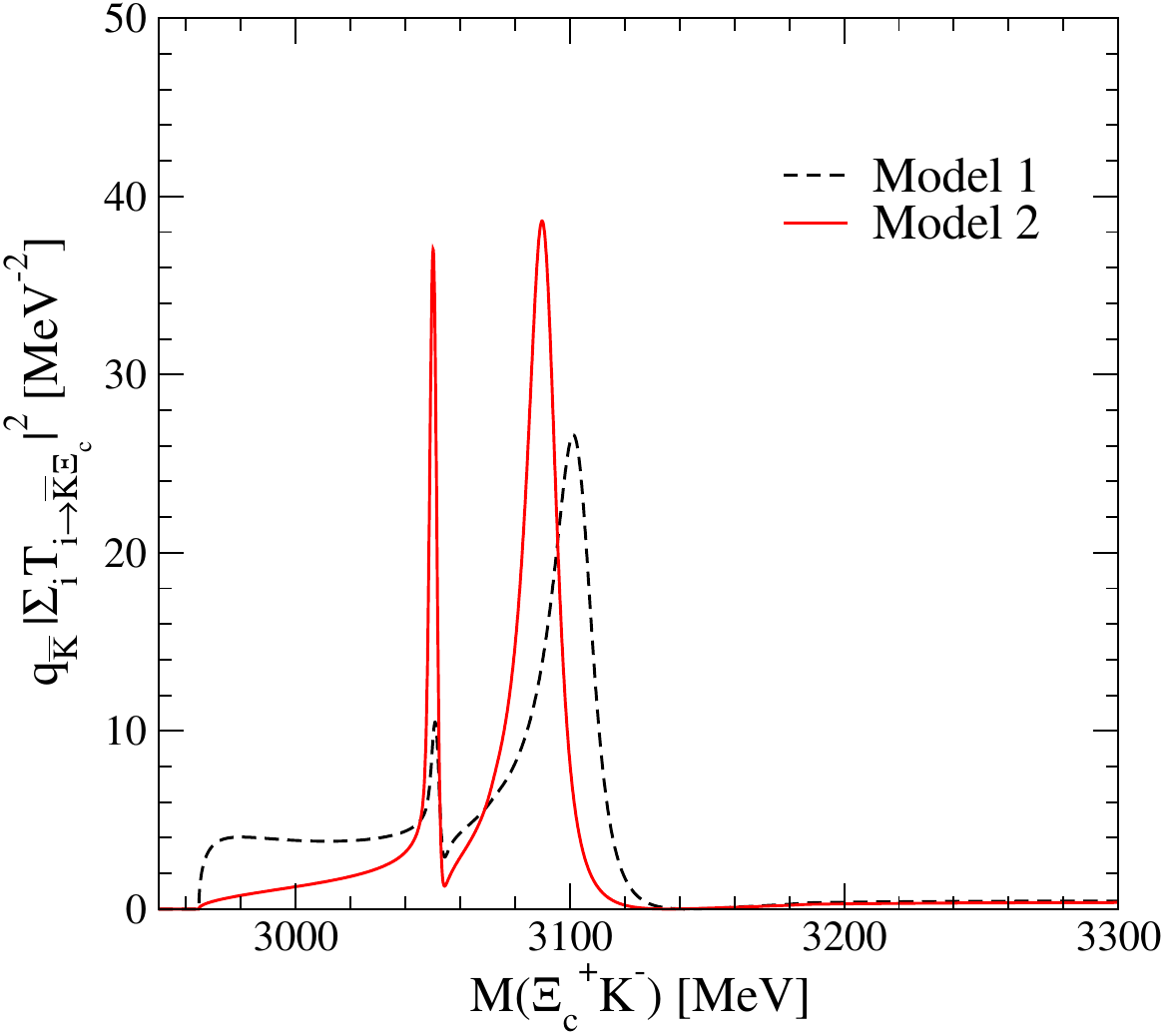}
\caption{(Color online) Sum of amplitudes squared times a phase space factor.} 
  \label{fig:t2}
\end{figure}

The five $\Omega_c^0$ states were observed from the $K^-\Xi^+_c$ invariant mass spectrum obtained from a sample of $pp$ collision data at center of mass energies of 7, 8 and 13 TeV, recorded by the LHCb experiment \cite{Aaij:2017nav}. To model such spectrum from the elementary $pp$ collision reaction is a tremendously difficult task, but we can give a taste of the spectrum that our models would predict by representing, in Fig.~\ref{fig:t2}, the quantity
\begin{equation}
q_{K^-} \mid \sum_{i} T_{i\to \bar{K}\Xi_c} \mid^2
\label{eq:t2}
\end{equation}
versus the $\bar{K}\Xi_c$ center-of-mass energy, where $T_{i\to \bar{K}\Xi_c}$ is the amplitude for the $i \to \bar{K}\Xi_c$ transition obtained here with either ``Model 1" (black dashed line) or ``Model 2" (red solid line), with $i$ being any of the five coupled channels involved in this sector. The momentum of the $K^-$ in the $\bar{K}\Xi_c$ center-of-mass frame, $q_{K^-}$, acts as a phase-space modulator. The calculated spectrum has been convoluted with the energy dependent resolution of the experiment, which runs linearly from 0.75~MeV at 3000~MeV to
1.74~MeV at 3119~MeV, employing a Gaussian function. We note that, in front of each amplitude $T_{i\to \bar{K}\Xi_c}$ in Eq.~ (\ref{eq:t2}), one should have included a coefficient gauging the strength with which the  production mechanism excites the particular meson-baryon channel $i$. Given the limited understanding of the production dynamics, we have assumed all these coefficients to be equal. Therefore, the spectrum displayed in Fig.~\ref{fig:t2} is merely orientative as it also lacks the background contributions. However, one can still see certain similarities with the spectrum of Fig.~2 in Ref.~\cite{Aaij:2017nav} in the energy regions of the 3050~MeV and 3090~MeV states. 

Finally, we construct the unitarized interaction between vector mesons and baryons in this sector, employing the set of coupling constants of Table~\ref{tab:a_vector}, which have been obtained  for a regularization scale of $\mu=1000$~GeV and imposing the loop function of each vector-baryon channel to coincide, at the corresponding threshold, with the cut-off loop function evaluated for $\Lambda=800$~MeV. The mass and other properties of the resonances found from the vector-baryon interaction in the $S=-2$, $C=1$ and $I=0$ sector are listed in Table~\ref{tab:vector}. We see a similar pattern as that found for the pseudoscalar-baryon case, one resonance coupling strongly to $D^*\Xi$ and the other coupling strongly to $\bar{K}^*\Xi'_c $ and to $\phi \Omega_c^0$, which mainly takes the role of the $\eta \Omega_c^0$ state of the pseudoscalar case according to the tranformation of Eq.~(\ref{eq:eta_phi}). However, the ordering in energies of these resonances appears interchanged with respect to that found in pseudoscalar-baryon scattering, which is simply related to the fact that the energy thresholds of the various vector-meson states have also changed with respect to their pseudoscalar-baryon counterparts. The lower energy resonance at 3231~MeV is mainly a $D^*\Xi$ bound state, while the resonance at  3419~MeV, is mainly a $\bar{K}^*\Xi'_c $ composite state with some admixture of  $\omega \Omega_c^0$ and
$\phi \Omega_c$ components. These resonances are located at energy values well above the states found by the LHCb collaboration in a region where no narrow structures have been seen \cite{Aaij:2017nav}. We note, however, that the states found here from the vector-baryon interaction are artificially narrow as they do not couple to, and hence cannot decay into, the pseudoscalar-baryon states that lie at lower energy. In order to account for this possibility in our model one should incorporate the coupling of vector-baryon states to the pseudoscalar-baryon ones, via e.g. box diagrams \cite{Garzon:2012np,Liang:2014kra} or employing the methodology of Refs.~\cite{Romanets:2012hm,GarciaRecio:2008dp} where, on the basis of heavy-quark spin symmetry, the pseudoscalar and vector mesons, as well as the baryons of the octet and those of the decuplet, are treated on the same footing. It would be interesting to perform such calculations in order to see if these structures remain narrow or widen up sufficiently to accommodate to the apparently featureless spectrum (within experimental errors)  in this higher energy range.
It would also be interesting to explore how the pseudoscalar-baryon resonances studied in the present work  would be affected by considering the coupling to the vector-baryon states, a task that goes beyond the scope of the present exploratory study. Note, however, that the energy threshold of the lighter $D^*\Xi$ vector-baryon channel lies above those of the pseudoscalar-meson channels, except for the $\eta^\prime \Omega_c^0$ one which plays a quite irrelevant role in the pseudoscalar-baryon states found here. We therefore expect limited changes in their energy positions and widths, which could anyway be compensated by appropriate changes in the subtraction constants.

\begin{table}[h]
\centering
\begin{tabular}{lccccc}
\hline 
\hline 
    &  $a_{ D^*\Xi}$  &  $a_{\bar{K}^*\Xi_c}$  &  $a_{\bar{K}^*\Xi'_c}$  &  $a_{\omega \Omega_c}$  & $a_{\phi \Omega_c}$ \\
\hline \\[-2.5mm]                             
  &  $-1.97$  &  $-2.15$  &  $-2.20$  & $-2.27$  &  $-2.26$   \\
\hline
\hline
\end{tabular}
\caption{{\small  Values of the subtraction constants for Model 1 at a regularization scale $\mu=1$ GeV.}} 
\label{tab:a_vector}
\end{table}

\begin{table}[h]
\centering
\begin{tabular}{c|cc|cc}
\hline
\hline
\multicolumn{5}{c}{ {\bf $1^- \oplus \frac{1}{2}^+$} interaction in {\bf$(I,S,C)=(0,-2,1)$} sector } \\ 
\hline \\[-2.5mm]
$M\;\rm[MeV]$             &    \multicolumn{2}{c|}{$3231.19$}    &   \multicolumn{2}{c}{$3419.25$}  \\
$\Gamma\;\rm[MeV]$   &     \multicolumn{2}{c|}{$0.0$}         &     \multicolumn{2}{c}{$4.8$}   \\ \hline \\[-2.5mm]
                                &   $| g_i|$   & $-g_i^2 dG/dE$        &   $| g_i|$   & $-g_i^2 dG/dE$     \\
$D^*\Xi (3326)$          &  $4.30$  &           $0.90-i0.00$      &   $0.24$  & $0.00+i0.00$      \\
$\bar{K}^*\Xi_c (3363)$  &  $0.64$  & $0.03-i0.00$    &   $0.13$  & $0.00+i0.00$      \\
$\bar{K}^*\Xi'_c (3470)$ &  $0.26$  & $0.00-i0.00$    &   $1.83$  &             $0.42+i0.02$      \\
$\omega \Omega_c (3480)$ &  $0.34$  & $0.01-i0.00$    &   $1.56$  &             $0.28+i0.00$      \\
$\phi \Omega_c (3717)$   &  $0.00$  & $0.00-i0.00$   &   $2.31$  &             $0.22+i0.00$      \\
\hline
\hline
\end{tabular}
\caption{
{\small  $\Omega^0_c (X)$ states generated employing a vector-type Weinberg-Tomozawa zero-range interaction
between a vector meson and a ground state baryon within a coupled-channel approach.}}
\label{tab:vector}
\end{table}

\section{Conclusions}\label{sec:conclusions}

In this work we have studied the interaction of the low-lying pseudoscalar mesons with the ground-state baryons in the charm $+1$, strangeness $-2$ and isospin $0$ sector, employing a  t-channel vector meson exchange model with effective Lagrangians.  
We unitarize the amplitude by means of the coupled-channel Bethe-Salpeter equation, paying a especial attention to regulate the loops with naturally sized subtraction constants. 

The resulting amplitude for the scattering of pseudoscalar mesons with baryons shows the presence of two resonances, having energies and widths very similar to some of the $\Omega_c^0$ states discovered recently at LHCb. By exploring the parameter space of our model we find several cases that can reproduce the mass and width of the $\Omega_c(3050)^0$ and  the $\Omega_c(3090)^0$.

Our findings allow us to conclude that two of the five $\Omega_c^0$ states recently observed by the LHCb collaboration could have a meson-baryon molecular origin. The state at 3050~MeV would mostly have a $\bar{K}\Xi'_c $ component (around 50\%) with a 20\% mixture of $\eta \Omega_c$, while the one at 3090~MeV would be essentially a $D\Xi$ molecule with a 90\% strength.

As our model for the scattering of pseudoscalar mesons with baryons in s-wave generates resonances with spin-parity $J^P=1/2^-$, we would anticipate these to be the quantum numbers for the 3050~MeV and 3090~MeV $\Omega_c^0$ states, in contrast to the expectations from quark models which establish either $3/2^-$ or $5/2^-$ for their spin-parity. 

An experimental determination of the spin-parity of the $\Omega_c^0$ states would be extremely valuable to disentangle the $3q$ or meson-baryon nature of some of the $\Omega_c^0$ states observed at LHCb. It is also expected that further theoretical studies about the molecular interpretation of baryons in the $S=-2$, $C=1$, $I=0$ sector, including additional components as the ones considered here, can bring new light into this problem.

\section*{Acknowledgments}

This work is partly supported by the Spanish Ministerio de Economia y Competitividad (MINECO) under the project MDM-2014-0369 of ICCUB (Unidad de Excelencia 'Mar\'\i a de Maeztu'), 
and, with additional European FEDER funds, under the contract FIS2014-54762-P.
Support has also been received from the Ge\-ne\-ra\-li\-tat de Catalunya contract 2014SGR-401.



\begin{thebibliography}{999}

\bibitem{Aaij:2017nav} 
  R.~Aaij {\it et al.} [LHCb Collaboration],
  Phys.\ Rev.\ Lett.\  {\bf 118}, no. 18, 182001 (2017)
  doi:10.1103/PhysRevLett.118.182001
  [arXiv:1703.04639 [hep-ex]].
  
\bibitem{Karliner:2017kfm} 
  M.~Karliner and J.~L.~Rosner,
  Phys.\ Rev.\ D {\bf 95}, no. 11, 114012 (2017)
  doi:10.1103/PhysRevD.95.114012
  [arXiv:1703.07774 [hep-ph]].
 
   
\bibitem{Wang:2017vnc} 
  W.~Wang and R.~L.~Zhu,
  Phys.\ Rev.\ D {\bf 96}, no. 1, 014024 (2017)
  doi:10.1103/PhysRevD.96.014024
  [arXiv:1704.00179 [hep-ph]].
  
\bibitem{Wang:2017zjw} 
  Z.~G.~Wang,
  Eur.\ Phys.\ J.\ C {\bf 77}, no. 5, 325 (2017)
  doi:10.1140/epjc/s10052-017-4895-5
  [arXiv:1704.01854 [hep-ph]].
 
\bibitem{Chen:2017gnu} 
  B.~Chen and X.~Liu,
  arXiv:1704.02583 [hep-ph].
 
   
\bibitem{Padmanath:2017lng} 
  M.~Padmanath and N.~Mathur,
  Phys.\ Rev.\ Lett.\  {\bf 119}, no. 4, 042001 (2017)
  doi:10.1103/PhysRevLett.119.042001
  [arXiv:1704.00259 [hep-ph]].
  
 
\bibitem{Chen:2017sci} 
  H.~X.~Chen, Q.~Mao, W.~Chen, A.~Hosaka, X.~Liu and S.~L.~Zhu,
  Phys.\ Rev.\ D {\bf 95}, no. 9, 094008 (2017)
  doi:10.1103/PhysRevD.95.094008
  [arXiv:1703.07703 [hep-ph]].

\bibitem{Agaev:2017jyt} 
  S.~S.~Agaev, K.~Azizi and H.~Sundu,
  Europhys.\ Lett.\  {\bf 118}, no. 6, 61001 (2017)
  doi:10.1209/0295-5075/118/61001
  [arXiv:1703.07091 [hep-ph]].

    
\bibitem{Agaev:2017lip} 
  S.~S.~Agaev, K.~Azizi and H.~Sundu,
  Eur.\ Phys.\ J.\ C {\bf 77}, no. 6, 395 (2017)
  doi:10.1140/epjc/s10052-017-4953-z
  [arXiv:1704.04928 [hep-ph]].
 
\bibitem{Cheng:2017ove} 
  H.~Y.~Cheng and C.~W.~Chiang,
  Phys.\ Rev.\ D {\bf 95}, no. 9, 094018 (2017)
  doi:10.1103/PhysRevD.95.094018
  [arXiv:1704.00396 [hep-ph]].
  
\bibitem{Wang:2017hej} 
  K.~L.~Wang, L.~Y.~Xiao, X.~H.~Zhong and Q.~Zhao,
  Phys.\ Rev.\ D {\bf 95}, no. 11, 116010 (2017)
  doi:10.1103/PhysRevD.95.116010
  [arXiv:1703.09130 [hep-ph]].

  
\bibitem{Huang:2017dwn} 
  H.~Huang, J.~Ping and F.~Wang,
  arXiv:1704.01421 [hep-ph].

    
\bibitem{Yang:2017rpg} 
  G.~Yang and J.~Ping,
  arXiv:1703.08845 [hep-ph].
  
\bibitem{An:2017lwg} 
  C.~S.~An and H.~Chen,
  Phys.\ Rev.\ D {\bf 96}, no. 3, 034012 (2017)
  doi:10.1103/PhysRevD.96.034012
  [arXiv:1705.08571 [hep-ph]].
 
\bibitem{Kim:2017jpx} 
  H.~C.~Kim, M.~V.~Polyakov and M.~Praszałowicz,
  Phys.\ Rev.\ D {\bf 96}, no. 1, 014009 (2017)
  Addendum: [Phys.\ Rev.\ D {\bf 96}, no. 3, 039902 (2017)]
  doi:10.1103/PhysRevD.96.039902, 10.1103/PhysRevD.96.014009
  [arXiv:1704.04082 [hep-ph]].
    




\bibitem{Capstick:1986bm} 
  S.~Capstick and N.~Isgur,
  Phys.\ Rev.\ D {\bf 34}, 2809 (1986)
  [AIP Conf.\ Proc.\  {\bf 132}, 267 (1985)].
  doi:10.1103/PhysRevD.34.2809, 10.1063/1.35361
  
\bibitem{pdg} 
  C.~Patrignani {\it et al.} [Particle Data Group],
  Chin.\ Phys.\ C {\bf 40}, no. 10, 100001 (2016).
  doi:10.1088/1674-1137/40/10/100001
  
  
\bibitem{Maltman:1980er} 
  K.~Maltman and N.~Isgur,
  Phys.\ Rev.\ D {\bf 22}, 1701 (1980).
  doi:10.1103/PhysRevD.22.1701
  
\bibitem{Migura:2006ep} 
  S.~Migura, D.~Merten, B.~Metsch and H.~R.~Petry,
  Eur.\ Phys.\ J.\ A {\bf 28}, 41 (2006)
  doi:10.1140/epja/i2006-10017-9
  [hep-ph/0602153].
   
\bibitem{Roberts:2007ni} 
  W.~Roberts and M.~Pervin,
  Int.\ J.\ Mod.\ Phys.\ A {\bf 23}, 2817 (2008)
  doi:10.1142/S0217751X08041219
  [arXiv:0711.2492 [nucl-th]].
   
\bibitem{Valcarce:2008dr} 
  A.~Valcarce, H.~Garcilazo and J.~Vijande,
  Eur.\ Phys.\ J.\ A {\bf 37}, 217 (2008)
  doi:10.1140/epja/i2008-10616-4
  [arXiv:0807.2973 [hep-ph]].
  
\bibitem{Ebert:2011kk} 
  D.~Ebert, R.~N.~Faustov and V.~O.~Galkin,
  Phys.\ Rev.\ D {\bf 84}, 014025 (2011)
  doi:10.1103/PhysRevD.84.014025
  [arXiv:1105.0583 [hep-ph]].
  
\bibitem{Vijande:2013yxa} 
  J.~Vijande, A.~Valcarce, T.~F.~Carames and H.~Garcilazo,
  Nucl.\ Phys.\ A {\bf 914}, 472 (2013).
  doi:10.1016/j.nuclphysa.2012.12.096

\bibitem{Yoshida:2015tia} 
  T.~Yoshida, E.~Hiyama, A.~Hosaka, M.~Oka and K.~Sadato,
  Phys.\ Rev.\ D {\bf 92}, no. 11, 114029 (2015)
  doi:10.1103/PhysRevD.92.114029
  [arXiv:1510.01067 [hep-ph]].

\bibitem{Oller:2000fj} 
  J.~A.~Oller and U.~G.~Meissner,
  Phys.\ Lett.\ B {\bf 500}, 263 (2001)
  doi:10.1016/S0370-2693(01)00078-8
  [hep-ph/0011146].
  
\bibitem{Jido:2003cb} 
  D.~Jido, J.~A.~Oller, E.~Oset, A.~Ramos and U.~G.~Meissner,
  Nucl.\ Phys.\ A {\bf 725}, 181 (2003)
  doi:10.1016/S0375-9474(03)01598-7
  [nucl-th/0303062].

\bibitem{Hyodo:2011ur} 
  T.~Hyodo and D.~Jido,
  Prog.\ Part.\ Nucl.\ Phys.\  {\bf 67}, 55 (2012)
  doi:10.1016/j.ppnp.2011.07.002
  [arXiv:1104.4474 [nucl-th]].

 \bibitem{Magas:2005vu}
   V.~K.~Magas, E.~Oset and A.~Ramos,
   Phys.\ Rev.\ Lett.\  {\bf 95}, 052301 (2005)
   [hep-ph/0503043].

\bibitem{Thomas:1973uh} 
  D.~W.~Thomas, A.~Engler, H.~E.~Fisk and R.~W.~Kraemer,
  Nucl.\ Phys.\ B {\bf 56}, 15 (1973).
  doi:10.1016/0550-3213(73)90217-4

\bibitem{Prakhov:2004an} 
  S.~Prakhov {\it et al.}  [Crystall Ball Collaboration],
  Phys.\ Rev.\ C {\bf 70}, 034605 (2004).
  
\bibitem{Niiyama:2008rt} 
  M.~Niiyama {\it et al.},
  Phys.\ Rev.\ C {\bf 78}, 035202 (2008)
  doi:10.1103/PhysRevC.78.035202
  [arXiv:0805.4051 [hep-ex]].
  
\bibitem{Moriya:2013hwg} 
  K.~Moriya {\it et al.} [CLAS Collaboration],
  Phys.\ Rev.\ C {\bf 88}, 045201 (2013)
  Addendum: [Phys.\ Rev.\ C {\bf 88}, no. 4, 049902 (2013)]
  doi:10.1103/PhysRevC.88.049902, 10.1103/PhysRevC.88.045201
  [arXiv:1305.6776 [nucl-ex]].
  
\bibitem{Roca:2013av} 
  L.~Roca and E.~Oset,
  Phys.\ Rev.\ C {\bf 87}, no. 5, 055201 (2013)
  doi:10.1103/PhysRevC.87.055201
  [arXiv:1301.5741 [nucl-th]].
  
\bibitem{Mai:2014xna} 
  M.~Mai and U.~G.~Meißner,
  Eur.\ Phys.\ J.\ A {\bf 51}, no. 3, 30 (2015)
  doi:10.1140/epja/i2015-15030-3
  [arXiv:1411.7884 [hep-ph]].

\bibitem{Aaij:2015tga} 
  R.~Aaij {\it et al.} [LHCb Collaboration],
  Phys.\ Rev.\ Lett.\  {\bf 115}, 072001 (2015)
  doi:10.1103/PhysRevLett.115.072001
  [arXiv:1507.03414 [hep-ex]].
  
\bibitem{Wu:2010jy} 
  J.~J.~Wu, R.~Molina, E.~Oset and B.~S.~Zou,
  Phys.\ Rev.\ Lett.\  {\bf 105}, 232001 (2010)
  [arXiv:1007.0573 [nucl-th]].
  
\bibitem{Wu:2010vk} 
  J.~J.~Wu, R.~Molina, E.~Oset and B.~S.~Zou,
  Phys.\ Rev.\ C {\bf 84}, 015202 (2011)
  [arXiv:1011.2399 [nucl-th]].
  
  
\bibitem{Yang:2011wz} 
  Z.~C.~Yang, Z.~F.~Sun, J.~He, X.~Liu and S.~L.~Zhu,
  Chin.\ Phys.\ C {\bf 36}, 6 (2012)
  [arXiv:1105.2901 [hep-ph]].
  
\bibitem{Xiao:2013yca} 
  C.~W.~Xiao, J.~Nieves and E.~Oset,
  Phys.\ Rev.\ D {\bf 88}, 056012 (2013)
  [arXiv:1304.5368 [hep-ph]].
  
\bibitem{Karliner:2015ina} 
  M.~Karliner and J.~L.~Rosner,
  Phys.\ Rev.\ Lett.\  {\bf 115}, no. 12, 122001 (2015)
  [arXiv:1506.06386 [hep-ph]].
 
\bibitem{Chen:2015loa} 
  R.~Chen, X.~Liu, X.~Q.~Li and S.~L.~Zhu,
  Phys.\ Rev.\ Lett.\  {\bf 115}, no. 13, 132002 (2015)
  [arXiv:1507.03704 [hep-ph]].
 
\bibitem{Roca:2015dva} 
  L.~Roca, J.~Nieves and E.~Oset,
  Phys.\ Rev.\ D {\bf 92}, no. 9, 094003 (2015)
  [arXiv:1507.04249 [hep-ph]].
  
\bibitem{He:2015cea} 
  J.~He,
  arXiv:1507.05200 [hep-ph].
 
\bibitem{Meissner:2015mza} 
  U.~G.~Meißner and J.~A.~Oller,
  Phys.\ Lett.\ B {\bf 751}, 59 (2015)
  [arXiv:1507.07478 [hep-ph]].

\bibitem{Chen:2015sxa} 
  H.~X.~Chen, L.~S.~Geng, W.~H.~Liang, E.~Oset, E.~Wang and J.~J.~Xie,
  arXiv:1510.01803 [hep-ph].
  
\bibitem{Feijoo:2015kts} 
  A.~Feijoo, V.~K.~Magas, A.~Ramos and E.~Oset,
  Eur.\ Phys.\ J.\ C {\bf 76}, no. 8, 446 (2016)
  doi:10.1140/epjc/s10052-016-4302-7
  [arXiv:1512.08152 [hep-ph]].
   
\bibitem{Guo:2017jvc} 
  F.~K.~Guo, C.~Hanhart, U.~G.~Meißner, Q.~Wang, Q.~Zhao and B.~S.~Zou,
  arXiv:1705.00141 [hep-ph].
  
\bibitem{Hofmann:2005sw} 
  J.~Hofmann and M.~F.~M.~Lutz,
  Nucl.\ Phys.\ A {\bf 763}, 90 (2005)
  doi:10.1016/j.nuclphysa.2005.08.022
  [hep-ph/0507071].

\bibitem{JimenezTejero:2009vq} 
  C.~E.~Jimenez-Tejero, A.~Ramos and I.~Vidana,
  Phys.\ Rev.\ C {\bf 80}, 055206 (2009)
  doi:10.1103/PhysRevC.80.055206
  [arXiv:0907.5316 [hep-ph]].
  
\bibitem{Romanets:2012hm} 
  O.~Romanets, L.~Tolos, C.~Garcia-Recio, J.~Nieves, L.~L.~Salcedo and R.~G.~E.~Timmermans,
  Phys.\ Rev.\ D {\bf 85}, 114032 (2012)
  doi:10.1103/PhysRevD.85.114032
  [arXiv:1202.2239 [hep-ph]].

\bibitem{Kawarabayashi:1966kd} 
  K.~Kawarabayashi and M.~Suzuki,
  Phys.\ Rev.\ Lett.\  {\bf 16}, 255 (1966).
  doi:10.1103/PhysRevLett.16.255
  
\bibitem{Riazuddin:1966sw} 
  Riazuddin and Fayyazuddin,
  Phys.\ Rev.\  {\bf 147}, 1071 (1966).
  doi:10.1103/PhysRev.147.1071
  
\bibitem{Gamermann:2006nm} 
  D.~Gamermann, E.~Oset, D.~Strottman and M.~J.~Vicente Vacas,
  Phys.\ Rev.\ D {\bf 76}, 074016 (2007)
  doi:10.1103/PhysRevD.76.074016
  [hep-ph/0612179].
  
\bibitem{Mizutani:2006vq} 
  T.~Mizutani and A.~Ramos,
  Phys.\ Rev.\ C {\bf 74}, 065201 (2006)
  doi:10.1103/PhysRevC.74.065201
  [hep-ph/0607257].
  
\bibitem{Oset:2009vf} 
  E.~Oset and A.~Ramos,
  Eur.\ Phys.\ J.\ A {\bf 44}, 445 (2010)
  doi:10.1140/epja/i2010-10957-3
  [arXiv:0905.0973 [hep-ph]].
  
\bibitem{Roca:2005nm} 
  L.~Roca, E.~Oset and J.~Singh,
  Phys.\ Rev.\ D {\bf 72}, 014002 (2005)
  doi:10.1103/PhysRevD.72.014002
  [hep-ph/0503273].
  
\bibitem{Weinberg:1965zz} 
  S.~Weinberg,
  Phys.\ Rev.\  {\bf 137}, B672 (1965).
  doi:10.1103/PhysRev.137.B672
  
\bibitem{Gamermann:2009uq} 
  D.~Gamermann, J.~Nieves, E.~Oset and E.~Ruiz Arriola,
  Phys.\ Rev.\ D {\bf 81}, 014029 (2010)
  doi:10.1103/PhysRevD.81.014029
  [arXiv:0911.4407 [hep-ph]].
  
\bibitem{Hyodo:2011qc} 
  T.~Hyodo, D.~Jido and A.~Hosaka,
  Phys.\ Rev.\ C {\bf 85}, 015201 (2012)
  doi:10.1103/PhysRevC.85.015201
  [arXiv:1108.5524 [nucl-th]].
  
\bibitem{Aceti:2012dd} 
  F.~Aceti and E.~Oset,
  Phys.\ Rev.\ D {\bf 86}, 014012 (2012)
  doi:10.1103/PhysRevD.86.014012
  [arXiv:1202.4607 [hep-ph]].
  
\bibitem{Hyodo:2013nka} 
  T.~Hyodo,
  Int.\ J.\ Mod.\ Phys.\ A {\bf 28}, 1330045 (2013)
  doi:10.1142/S0217751X13300457
  [arXiv:1310.1176 [hep-ph]].

\bibitem{Aceti:2014ala} 
  F.~Aceti, L.~R.~Dai, L.~S.~Geng, E.~Oset and Y.~Zhang,
  Eur.\ Phys.\ J.\ A {\bf 50}, 57 (2014)
  doi:10.1140/epja/i2014-14057-2
  [arXiv:1301.2554 [hep-ph]].
  
  
\bibitem{Garzon:2012np} 
  E.~J.~Garzon and E.~Oset,
  Eur.\ Phys.\ J.\ A {\bf 48}, 5 (2012)
  doi:10.1140/epja/i2012-12005-x
  [arXiv:1201.3756 [hep-ph]].
  
\bibitem{Liang:2014kra} 
  W.~H.~Liang, T.~Uchino, C.~W.~Xiao and E.~Oset,
  Eur.\ Phys.\ J.\ A {\bf 51}, no. 2, 16 (2015)
  doi:10.1140/epja/i2015-15016-1
  [arXiv:1402.5293 [hep-ph]].

\bibitem{GarciaRecio:2008dp}
  C.~Garcia-Recio, V.~K.~Magas, T.~Mizutani, J.~Nieves, A.~Ramos, L.~L.~Salcedo and L.~Tolos,
  Phys.\ Rev.\ D {\bf 79} (2009) 054004
  doi:10.1103/PhysRevD.79.054004
  [arXiv:0807.2969 [hep-ph]].
  
  
  
\end{thebibliography}
\end{document}